# Haptic Color Patterns for Visually Impaired People
# -Pilot Study for a Learning Color Wheel-

Hsin-Yi Chao [1], and Hiroki Ishizuka[2]

[1] Program of Digital Humanities and Creative Industries & Graduate Institute of Library and Information Science, National Chung Hsing University, Taichung, Taiwan

[2] Graduate School of Engineering Science, Osaka University, Toyonaka, Japan

(Email: chy@nchu.edu.tw)

**Abstract ---** **This study proposes a tactile diagram pattern for visually impaired people to recognize color information. The pattern uses the principle of three primary colors, with different patterns representing red, blue, and yellow. The size of tactile elements on these patterns indicates the proportion of the color mixing. A preliminary experiment showed that even a sighted participant could understand and reconstruct the tactile diagram. Future experiments will target visually impaired people to confirm the effectiveness of this method.**

**Keywords: tactile diagram, accessibility, color education, tactile perception, visual impairment**

## 1. Introduction

We recognize color as visual information and utilize this information in our daily lives. For instance, we discern the type and state of objects from their color, and we receive information from our surroundings through color, such as traffic signals. Therefore, color is an essential element of information for humans to carry out our daily lives.

Despite the importance of color information in daily life, visually impaired people need help to utilize this information directly. There is a need to recognize color information in many aspects of daily life, such as coordinating clothes, checking the completion of cooking, and operating household appliances. Visually impaired people may feel inconvenienced when performing these activities. The inconvenience can lead to a decline in the quality of daily life for visually impaired people. Moreover, because visually impaired people cannot directly understand information based on these colors, their access to information is limited. As a result, visually impaired individuals are at risk of missing important information. Therefore, it is essential to establish a means for visually impaired individuals to utilize color information.

In this study, we propose a tactile diagram pattern for color, utilizing the principle of the three primary colors. A straight-line pattern represents red, a wavy-line pattern represents blue, and a dot pattern represents yellow. The size of these patterns represents the intensity of the primary color. The combination of the patterns and their size reproduces the color information. There have been studies that have attempted to reproduce color using braille patterns or tactile diagram patterns. In 2009, Filipa Nogueira Pires from the University of Lisbon and the Helen Keller Center developed the Feelipa Color Code, a tactile system for the visually impaired. Using geometric shapes based on Kandinsky's research—red as a square, yellow as a triangle, and blue as a circle—it forms 24 symbols for color identification. While effective for single objects, its design limits use in large-scale images [1]. The Scripor Alphabet, created by Tudor Scripor in 2019, is a braille-inspired tactile system using a 10-dot matrix. The reference dot above the second column aids in positioning and prevents confusion with braille. It represents nine colors—primary, secondary, and others—by combining dots. Although similar to braille, its lack of clear color associations and resemblance to braille text makes it difficult to distinguish between color symbols and text [2]. In addition to braille and shape patterns, there were problems such as different notations for each country and the need to read from a specific direction [3, 4]. In the case of existing tactile diagram patterns, there were problems, such as only designing patterns that represent specific colors and no logical rules existing between the patterns [5].

The tactile diagram pattern proposed in this study is a repetition of patterns such as lines, waves, and dots, so that it can be recognized from any direction [6]. Also, blending the patterns of the primary colors, just like actual colors, presents color information, making it possible to learn and understand actual color information. The proposed pattern can facilitate the use of color information by late visually impaired people and help early visually impaired individuals understand color information. In this study, we propose a system for learning color information using this tactile diagram and conduct a preliminary experiment.

## 2. DESIGN

Fig. 1 shows a color wheel using the proposed tactile diagram patterns. The straight line, wavy line, and dot patterns represent the primary colors of red, blue, and yellow, respectively. The design was determined by preliminary considerations that targeted visually impaired people. To facilitate the learning of the user, the ring is likened to a clock. Yellow is placed at 12 o'clock when the sun is high. Red is placed at 4 p.m., which is a hot time. Blue is placed at 8 o'clock, which is a cold time. At the intermediate positions of these colors, patterns of secondary colors, purple, green, and orange, which are mixed colors of each color, are placed. They are combinations of equal amounts of the patterns of the primary colors. Tertiary colors are placed between the primary and secondary colors. The mixing ratio of the primary colors determines the ratio of the size of the pattern. Actual patterns are shown in Fig. 2.

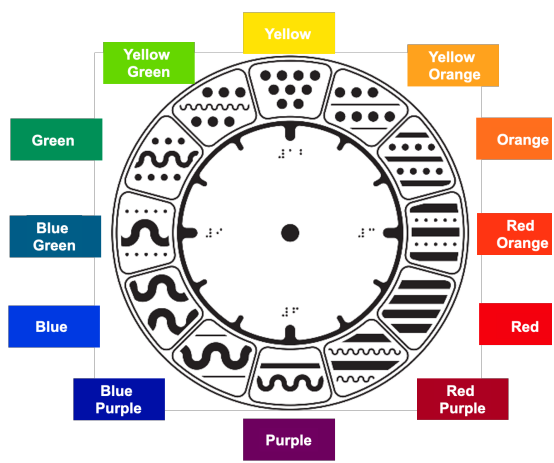

Fig. 1 Raised tactile diagram patterns for color wheel.

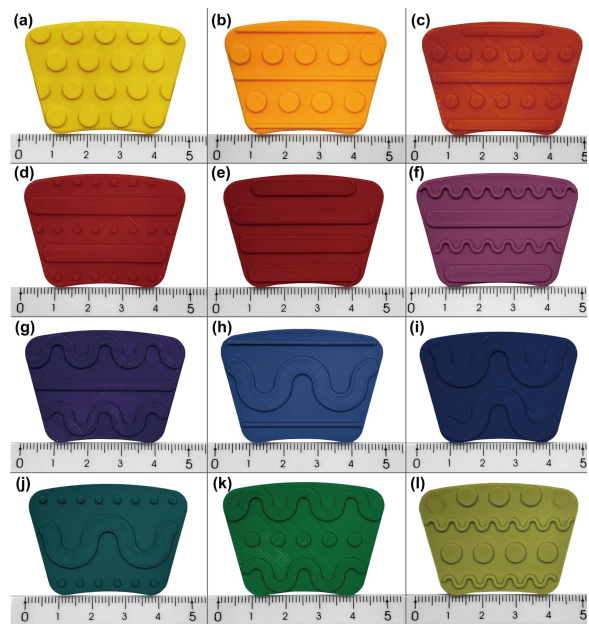

Fig. 2 Photographs of tactile diagram patterns. (a)Yellow. (b) Yellow orange. (c) Orange. (d) Red

The raised patterns are printed on 12 pieces with different size according to the proportion of primary colors. Secondary colors, orange, green and purple, are showed the equal amount pigment from two of three primary colors, red, yellow, and blue. So, the pattern sizes of dot and line are similar.

However, the six tertiary colors show the different size to stand for various ratio of primary colors. Yellow orange is presented as the larger size of dot stand for more yellow and the thinner straight lines stand for less red. Red orange is presented as the smaller size of dot stand for less yellow and the thicker straight lines stand for more red. Yellow green is presented as the larger size of dot stand for more yellow and the thinner wavy lines stand for less blue. Blue green is presented as the smaller size of dot stand for less yellow and the thicker wavy lines stand for more blue. Red purple is presented as the thicker straight lines stand for more red and the thinner wavy lines stand for less blue. Blue purple is presented as as the thinner straight lines stand for less red and the thicker wavy lines stand for more blue.

These logical designs of color wheel increase the abilities of comprehend the rule of color mixing and association for the visually impaired people

## 3. EXPERIMENT

### 3.1 Experimental Procedure

We conducted a preliminary experiment. The experimental participant was a sighted male in his 20s, wearing an eye mask for the experiment (Fig. 3(a)).

In the experiment, the participant first touched the fixed color wheel and heard the explanation for the relationship between color information and the tactile diagram pattern while touching it (Fig. 3(b)).

We explained the positions of the primary colors and how the secondary colors and then the tertiary colors are determined by their combinations. We advised him to start placing the pieces from the primary colors for the reconfiguration. When the participant said that he understood, we had him perform the task of reconfiguring the pattern. We asked him to place the 12 pieces formed from the patterns for the 12 colors in the appropriate positions in a case (Fig. 3(c)). The participant worked until he judged that he had completed it.

During the whole process of the task, we recorded the video from the side and top view to analyses how the participant utilizes the logical strategy of representing the correct arrangement of color wheel.

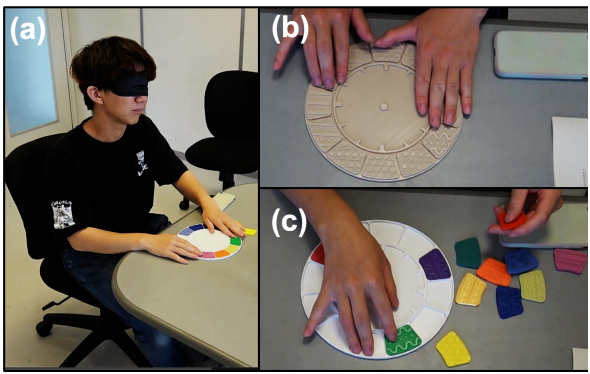

Fig. 3 Photographs of an experiment. (a) A participant attended wearing a m eye mask. (b) Firstly, he learned the principle of tactile diagram patters for color. (c) Then, he was asked to places in the appropriate positions.

### 3.2 Result & Discussion

Fig. 4 shows the color wheel reconstructed by the participant and the correct answer. It took six and a half minutes to reconstruct, and he was able to place four pieces in the appropriate positions. After placing the red piece of the primary color, he placed the purple piece in the place where the blue piece should have been placed. This is presumed to be because the participant could not correctly identify the patterns' shape. After placing several pieces, he placed the yellow piece of the primary color in the correct position. From this, he may remember the relationship between the patterns and the positions. After that, he placed the remaining pieces in the case. Through this experiment, we confirmed that even the sighted participant can understand the principles and arrangement of the tactile diagram and reconstruct it.

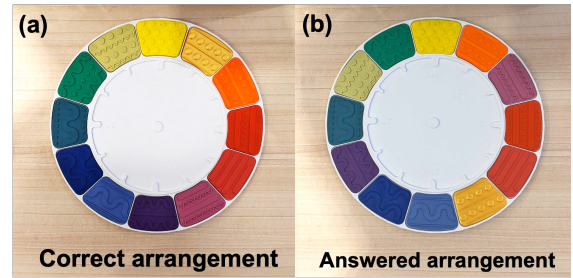

Fig. 4 The experimental results. (a) Correct arrangement for the reference. (b) Answered arrangement by the participant.

In future experiments targeting visually impaired people, the following can be inferred. Since visually impaired people who lose vision from birth may use touch more frequently than sighted individuals, their tactile perception is thought to be more sensitive. In addition, more touch experience may increase the ability to recognize the differences between tactile diagram. Therefore, it is thought that they will represent the tactile color pieces correctly. Also, they can effectively learn the logical principles of the tactile diagram through touch.

As a result, it is expected that they will be able to reproduce the color wheel as instructed. We want to clarify these points by conducting experiments targeting people with congenital and acquired visually impairments and blindness in the future.

## 4. CONCLUSION

In this paper, we propose a tactile diagram pattern for recognizing the color information of visually impaired people. Then, we design a color wheel using it and confirm whether users can learn the arrangement of the pattern through it. Through the experiment, the user was able to arrange it to some extent. In the future, we will conduct experiments targeting visually impaired individuals and confirm the usefulness of the proposed method.

ACKNOWLEDGEMENT

This work was supported in part by NTSC Grant Number 111-2410-H-005 -048 -MY2 of Taiwan and JST PRESTO Grant Number JPMJPR22S2 of Japan.